\documentclass[fleqn,twoside]{article} 
\usepackage{espcrc2}
\usepackage{graphicx}
\title{On the effective Hamiltonian for QCD: An overview and status report}
\author{Hans-Christian Pauli 
        \address{Max-Planck-Institut f\"ur Kernphysik, 
                 D-69029 Heidelberg, Germany}} 
\begin{document}
\begin{abstract}
    The session on effective Hamiltonians and chiral dynamics is 
    overviewed, combined with a review on the bound-state problem. 
    The progress during this session allows to remove all dependence 
    on regularization in an effective interaction, thus to renormalize
    a Hamiltonian for the first time, and to solve front form as if they 
    were instant-form equations, with all the advantages implied.
    \hfill                       HCP              4/11/20 November 2001 
\vspace{1pc}
\end{abstract}
\maketitle
\tableofcontents

\section{Introduction}
This community has formed in 1991 at the first light-cone meeting 
in Heidelberg with the ambitious aim to solve 
the bound-state problem in gauge field theory particularly QCD.

How far did we get?
Is it fair to say that we have not yet solved 
our homework problem?

However, the important contributions at this 
\cite{Leutwyler01,Roberts01,FredericoFP01,Schweiger01,Plessas01,Krassnigg01,Mangin01,FrewerPF01,Walhout01,Karmanov01,Ligterink01,Sugihara01,vanIersel01}
and the last meeting \cite{Ash00,Wegner00,Hil00,Schierholz00,TriPau00,Pau00}
let expect a faster pace in the overseeable future.
Over 80 participants show how alive the field continues
to be even after 11 years.
Its richness has become apparent last week.
13 speakers are alone in this session.
I attempt therefore to combine an overview on the present session on
``Effective Hamiltonians and chiral dynamics''
with a review on effective interactions in general. 
This seems to be in place in view of the progress at this meeting, 
particularly on renormalization \cite{FredericoFP01,FrewerPF01}
and the possibility to solve instant form rather than front form 
equations \cite{Krassnigg01} when working on the light cone.

\section{Why working on the light-cone?}
The Hamiltonian approach to a field theory was a no-go-topic
for over fifty years.
But combined with light-cone quantization and periodic boundary
conditions \cite{pab85a}, certain adventages 
inherent to the light-cone Hamiltonian approach 
were clear right from the outset, particularly 
\begin{itemize}
   \item  the simple kinematical boosts   and
   \item  the simple vacuum properties.  
\end{itemize}
This continues to be so.
It was equally clear that a number of extremely difficult
problems were on the road, among them
   zero modes,
   gauge invariance and gauge artifacts,
   the field theoretical many-body problem and
   Fock space truncation,
   non-perturbative renormalization,
   confinement,
   chiral phase transitions,
just to name a few.
Some of them have been solved, or better understood, 
as reviewed in \cite{BroPauPin98}.

What is the homework problem?
Starting from the Lagrangian density $\mathcal{L}_\mathit{QCD}$, 
the light-cone approach to the bound-state problem\cite{BroPauPin98}
aims at solving the eigenvalue equation
\begin{equation}  
   H_{LC}\vert\Psi\rangle = M^2\vert\Psi\rangle
.\end{equation}  
If one disregards possible zero modes 
and works in the light-cone gauge, 
the (light-cone) Hamiltonian $H_{LC}=P^\mu P_\mu$ 
is a well defined Fock-space operator and given in \cite{BroPauPin98}. 
Its eigenvalues are the invariant mass-squares  $M^2$
of physical particles associated with the eigenstates
$\vert\Psi\rangle$.
In general, they are superpositions of all possible
Fock states with its many-particle configurations.
For a meson, for example, holds 
\begin{center}
\(\displaystyle  
\begin{array} {rcll} 
      {\vert\Psi_{\rm meson}\rangle} &=& {\sum\limits_{i}\ }
      {\Psi_{q\bar q}(x_i,\vec k_{\!\perp_i},\lambda_i)}
      &{\vert q\bar q\rangle} 
\\&+& {\sum\limits_{i}\ }
      {\Psi_{g g}(x_i,\vec k_{\!\perp_i},\lambda_i)}
      &{\vert g g\rangle }
\\&+& {\sum\limits_{i}\ }
      {\Psi_{q\bar q g}(x_i,\vec k_{\!\perp_i},\lambda_i)}
      &{\vert q\bar q g\rangle }
\\&+&  {\sum\limits_{i}\ }
      {\Psi_{q\bar q q\bar q }(x_i,\vec k_{\!\perp_i},\lambda_i)}
      &{\vert q\bar q q\bar q \rangle}
\\&+& {\dots}.
\end{array}\)
\end{center}
If all wave functions $\Psi_n(x_i,\vec k_{\!\perp_i},\lambda_i)$
are available, one can analyze hadronic structure 
in terms of quarks and gluons.
For example, one can calculate the space-like form factor 
of a hadron quite straightforwardly by a
sum of overlap integrals analogous to the corresponding
non-relativistic formula \cite{BroPauPin98}.

\section{What are possible alternatives?}
This community tries hard to have a feedback to and from other fields 
and activities.
All these approaches have their own virtues and merits. 
The adventages are usually emphasized by the proponents,
and therefore I shall play the devils advocate passing them shortly review,
with a sometimes over-critical attitude to make the point clear.

\emph{Phenomenological models}.
Practically all our knowledge on hadron structure
comes from phenomenological models.
The constituent quark model particularly continues to have great success. 
Phenomenological approaches usually do not address to the lighter mesons
like the pion, but they are extremely succesfull for the heavier hadrons
and for baryons. A particularly beautiful example was presented by
Plessas \cite{Plessas01}. 
I do not care so much that his model has about twenty parameters,
with only three of them fitted explicitly.
Such work is a useful guideline to experiment.
I dream of the day when front-form based work produces similar
results. His work also shows the extreme difficulty
to relate wave functions of constituents to actual cross sections.\\
\emph{Chiral perturbation theory}.
Leutwyler \cite{Leutwyler01} demonstrates to which precision
a well based formalism can be driven. 
To some extent this also holds for the similar NJL-models. 
I cannot quote the huge body of literature 
but I mention in passing that they are not renormalizable,
that the relation to QCD is unclear, 
and that they deal mostly with the very light mesons. 
Heavy flavors cannot be treated, see also \cite{Leutwyler01}.\\
\emph{Schwinger-Dyson approaches} are potentially able 
to cope with the bound-state problem. 
Roberts \cite{Roberts01} emphasises the chiral aspects:
Free quarks have the small current mass at large momentum, 
increasing to the large constituent mass at small momentum.
Does this feature prevail in a bound state
problem, and how?\\
\emph{DLCQ and LC approaches}.
Hiller  \cite{Hil00} addresses to diagonalize by DLCQ 
the light-cone Hamiltonian 
in physical space-time (3+1).
His renormalization \`a la Pauli-Villars yields
promising results, but he needs a super-computer to produce them. 
He works in a truncated Fock space.
But Ligterink \cite{Ligterink01} concludes that Fock-space suppression
is less dangerous than believed,
and both Mangin-Brinet \cite{Mangin01} and Karmanov \cite{Karmanov01} 
report good stability of bound-state calculations in truncated spaces.
The separation of soft and hard aspects by Schweiger \cite{Schweiger01}
continue to be an important aspect of light-cone quantization.\\
\emph{Technical problems}.
Basis optimalization by Sugihara \cite{Sugihara01}
and a new algorithm by van Iersel \cite{vanIersel01},
applied here to the Yukawa model, are very important facets.
In fact the break-through
in non-perturbative renormalization by 
Frederico \cite{FredericoFP01} and Frewer \cite{FrewerPF01}, 
and a new insight into the nature of Melosh-transforms 
by Krassnigg \cite{Krassnigg01}
represent progress on a technical level as well.

\emph{Lattice Gauge Calculations} use practically all computer power 
in this world to generate a potential energy
between quarks, but then a \emph{non-relativistic}
Schr\"odinger equation is used to calculate bound states
\cite{Schierholz00}. This is unsatisfactory.
It is generally not known that LGC's have considerable
uncertainty to extrapolate their results down to such
light mesons as a pion.
It is equally unknown that lattice gauge calculations
get \emph{ always strict and linear confinement}
even for QED, where we know the ionization threshold.
The `breaking of the string', or in a more physical language,
the ionization threshold is one of the hot topics
at the lattice conferences \cite{Schilling2000}.
Moreover, in order to get the size of the pion,
thus the form factor, another generation of computers 
is required, as well as physicists to run them. 
Such considerations and the lacking perspectives on precision
have motivated Wilson, among other, to quit.
 
\emph{The new Wilson approach} to QCD is based almost entirely
on the front form and  renormalization group analysis \cite{Wilson6}.
Walhout \cite{Walhout01} gives an example for that.
The original hope was to assemble the operators 
in an effective interaction according to their relevance 
with respect to the renormalization group.
It is not unfair to state that not much of concrete hardcore technology 
has thus far emerged, despite the immense efforts over the years.
In developing the formalism the similarity transform of Wilson and
Glazek has played a major role \cite{GlazekWilson12}.
The basic idea is similar as in the preceding method 
\emph{Hamiltonian flow} by Wegner \cite{Weg94}.
But as emphasized repeatedly by the latter, 
the similarity transform has a serious defect \cite{Wegner00}:
As a built-in feature, it cannot account
for the block structures in
a gauge-field theoretical many-body Hamiltonian and
therefore should be abandoned. 

In conclusion, I believe that there is not much left than to proceed
with the more conventional methods.
In the sequel, I will briefly review only one of them, 
the method of iterated resolvents.

\begin{figure} [t]
  \scalebox{0.46}{\includegraphics{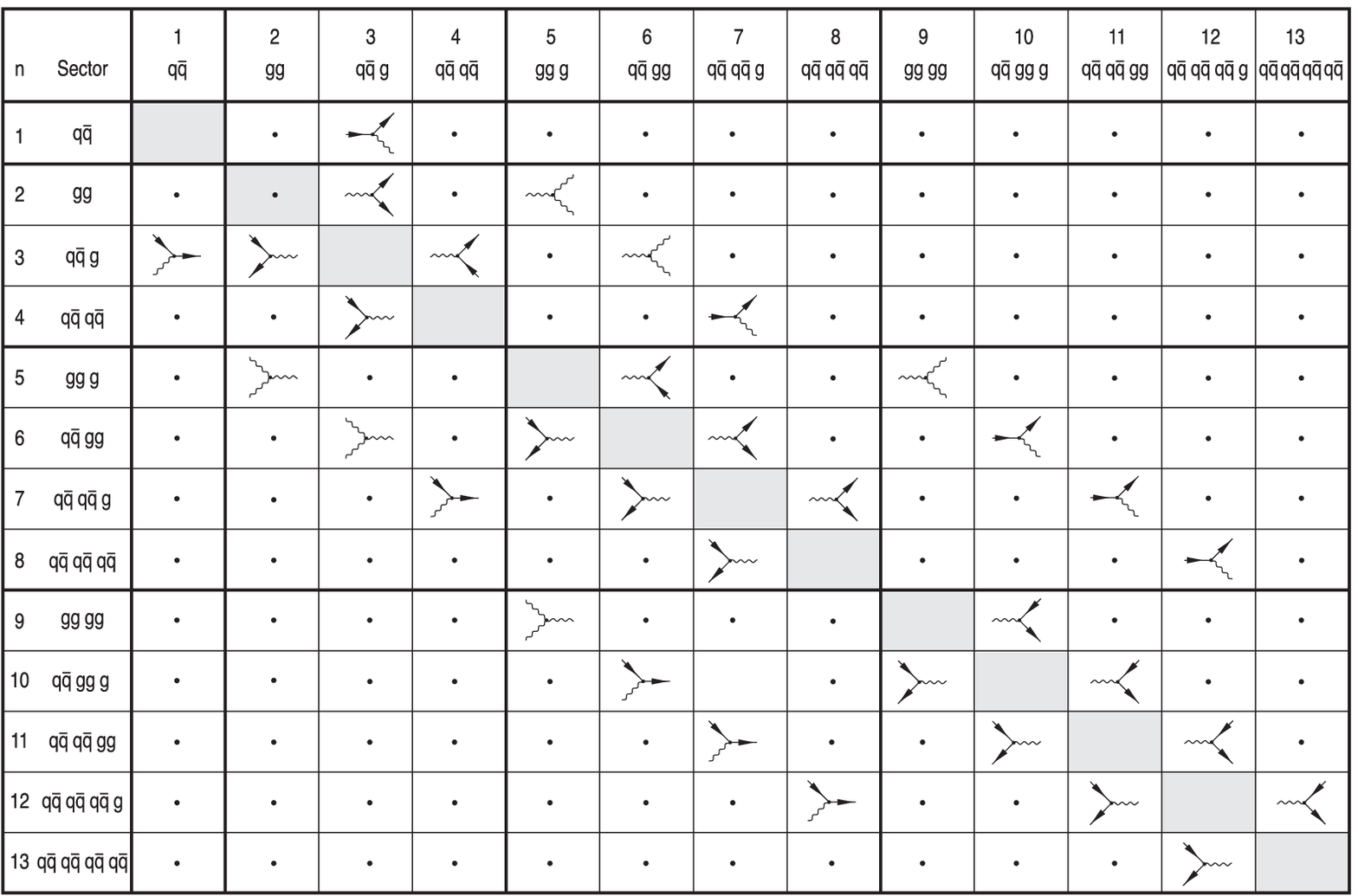}}
  \caption{\label{fig:1}  
     The Hamiltonian matrix for a meson. 
     The matrix elements are represented by energy diagrams. 
     Only vertex diagrams $V$ are shown.
     Zero matrices are marked by a dot ($\cdot$).} 
\end{figure}

%
\section {The method of iterated resolvents}
Instead of diagonalizing the Hamiltonian by DLCQ, 
one might wish to reduce the many-body problem
behind a field theory to an effective one-body problem.
The derivation of the effective interaction becomes then the key issue.

Because of the inherent divergencies in a gauge field theory,
the QCD-Hamiltonian in 3+1 dimensions must be regulated
from the outset. 
One of the few practical ways is vertex 
regularization \cite{BroPauPin98,Pau98},
where every Hamiltonian matrix element, particularly those of the 
vertex interaction (the Dirac interaction proper), 
is multiplied with a convergence-enforcing momentum-dependent function.
It can be viewed as a form factor \cite{BroPauPin98}.
The precise form of this function is unimportant here,
as long as it is a function of a cut-off scale ($\Lambda$).

By definition, an effective Hamiltonian acts only
in the lowest sector of the theory  
(here: in the Fock space of one quark and one anti-quark). 
And, again by definition, it has the same eigenvalue spectrum
as the full problem.
I have derived such an effective interaction 
by the method of iterated resolvents \cite{Pau98}, that is
by systematically expressing the higher Fock-space
wave functions as functionals of the lower ones.  
In doing so the Fock-space is not truncated
and all Lagrangian symmetries are preserved.
The projections of the eigenstates onto the 
higher Fock spaces can be retrieved
systematically from the $q\bar q$-projection, 
with explicit formulas given in \cite{Pau98}.

\begin{figure} [t]
\scalebox{0.44}{\includegraphics{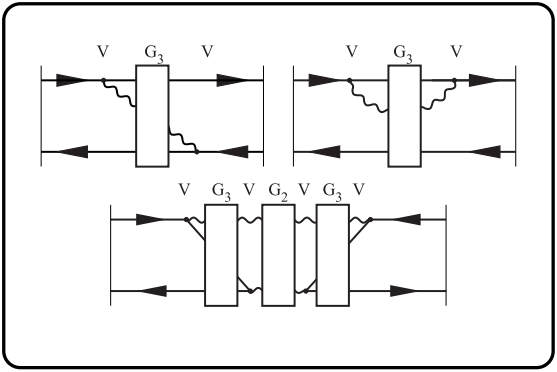}}
\caption{\label{fig:2}%
   The dressed propagators.}
\end{figure}

Let me sketch the method briefly, 
details may be found in \cite{Pau98}.
DLCQ with its periodic boundary conditions has the advantage
that the LC-Hamiltonian is a matrix with a finite number of
Fock-space sectors, which we denumerate by $n$, 
with $1 < n\leq N$.  
The so called harmonic resolution $K= L P^+/(2\pi)$
acts as a natural cut-off of the particle number.
As shown in Figure~\ref{fig:1}, $K=3$ allows for $N=8$, 
and $K=4$ for $N=13$ Fock-space sectors, for example.
The Hamiltonian matrix is sparse: Most of the
matrix elements are zero,
particularly if one includes only the vertex interaction $V$.
For $n$ sectors, the eigenvalue problem in terms of block matrices reads 
\begin{eqnarray} 
   \sum _{j=1} ^{n} \langle i \vert H _n (\omega)\vert j \rangle 
                    \langle j \vert\Psi  (\omega)\rangle 
   =  E (\omega)\ \langle i \vert\Psi (\omega)\rangle 
,\label{eq:2}\end{eqnarray} 
for $i=1,2,\dots,n$.
I can always invert the quadratic block matrix of the Hamitonian in 
the last sector to define the $n$-space resolvent $G _ n$, that is
\begin{eqnarray} 
    G _ n (\omega) =   {1\over \omega- H_n (\omega)} 
.\end{eqnarray}
Using $G _ n$, I can express the projection of the eigenfunction 
in the last sector by 
\begin{eqnarray} 
   \langle n \vert \Psi (\omega)\rangle   
   = G _ n (\omega) 
   \sum _{j=1} ^{n-1} \langle n \vert H _n (\omega)\vert j \rangle 
   \ \langle j \vert \Psi (\omega) \rangle 
,\end{eqnarray}
and substitute it in Eq.(\ref{eq:2}).
I then get an effective Hamiltonian where the number is sectors
is diminuished by 1:
\begin {equation}  
       H _{n -1} (\omega) =  H _n (\omega)
  +  H _n(\omega) G _ n  (\omega) H _n (\omega)
.\end {equation}
This is a recursion relation, which can repeated until one arrives
at the $q\bar q$-space.
The fixed point equation  $ E  (\omega ) = \omega $ determines 
all eigenvalues.

\begin{figure} [t]
\scalebox{0.44}{\includegraphics{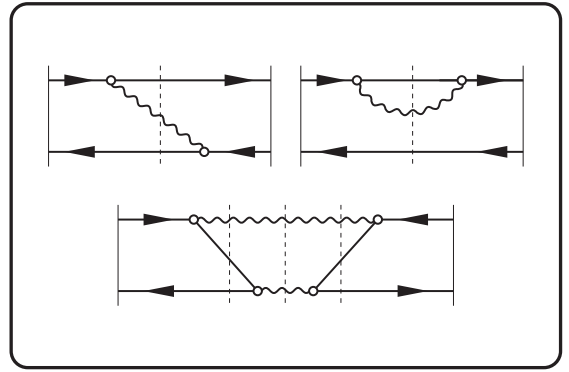}}
\caption{\label{fig:3}%
   The free propagators.}
\end{figure}

For the block matrix structure as in Figure~\ref{fig:1}, with its many 
zero matrices, the reduction is particularly easy and transparent.
For $K=3$ one has a sequence of effective interactions:
\begin{eqnarray}
\begin{array} {@{}l@{\,}c@{\,} l l@{\,}c@{\,}l@{}l@{}}
    H_8 &=& T_{8} ,  
  & H_7 &=& T_{7} + V G _8 V , 
\\  
    H_6 &=& T_{6} + V G _7 V , 
  & H_5 &=& T_{5} + V G _6 V.
\end{array}
\label{eq:6}\end{eqnarray}
The remaining ones get more complicated, \textit{i.e.}
\begin{eqnarray}
\begin{array} {@{}l@{\,}c@{\,}l@{}}
    H_4 &=& T_{4} + V G _7 V + V G _7 V  G _6 V G _7 V 
,\\  
    H_3 &=& T_{3} + V G _6 V + V G _6 V  G _5 V G _6 V + V G _4 V 
,\\  
    H_2 &=& T_{2} + V G _3 V + V G _5 V 
,\\  
    H_1 &=& T_{1} +  V G _3 V + V G _3 V  G _2 V G _3 V .
\end{array}
\hspace{-5em}\label{eq:7}\end{eqnarray}
For $K=4$, the effective interactions in Eq.(\ref{eq:6}) are different, 
see for example \cite{Pau98}, but it is quite remarkable that 
they are the same in Eq.(\ref{eq:7}).
In fact, the effective interactions in sectors 1-4 are independent of $K$:
The \emph{continuum limit $K\rightarrow\infty$ is then trivial},
and will be taken in the sequel. 

In the continuum limit, the effective Hamiltonian 
in the $q\bar q$-space $H_1=H_\mathit{eff}$ is thus
\begin{eqnarray}
\begin{array} {@{}l@{\,}c@{\,} l@{\,} l @{\,}l @{\,}l @{\,}l@{\,}}
    H_\mathit{eff} &=& T &+& V G _3 V &+& V G _3 V  G _2 V G _3 V ,  
\\  
                   &=& T &+& U_\mathit{conser} 
		         &+& U_\mathit{change} . 
\end{array}
\label{eq:8}\end{eqnarray}
The effective interaction has two contributions:
A flavor-conserving $U_\mathit{conser}$
and a flavor-changing piece $U_\mathit{change}$.
The flavor-changing interaction can not get active in flavor-off-diagonal mesons.

The dressed propagators in Eq.(\ref{eq:8}) and Figure~\ref{fig:2} are exact.
The iterated resolvents resum perturbative diagrams to all orders.

Their conversion to free propagators with effective vertices 
$\alpha\rightarrow\overline\alpha (Q)$,
represented in Figure~\ref{fig:3} by the thin lines and the open circles, 
respectively, is an approximation coupled with four well specified
assumptions \cite{Pau98}.

The wavy line in Figure~\ref{fig:3} should not be mistaken 
as a single gluon exchange.
The effective gluon corresponds to a particular resummation
of infinitely many gluons.

Henceforward I deal only with the flavor conserving interaction.
\section{The one-body equation}
The effective one-body equation 
for flavor off-diagonal mesons
(mesons with a different flavor for quark and anti-quark) 
becomes thus \cite{Pau00,Pau98}:
\begin{eqnarray} 
    M^2\psi_{\lambda_1\lambda_2}(x,\vec k_{\!\perp}) 
    = \sum _{ \lambda_1',\lambda_2'} 
    \!\!\int\!\! dx' d^2 \vec k_{\!\perp}' 
\nonumber\\ \times   
    \ U_{\lambda_1\lambda_2;\lambda_1'\lambda_2'}
    (x,\vec k_{\!\perp};x',\vec k_{\!\perp}')
    \ \psi_{\lambda_1'\lambda_2'}(x',\vec k_{\!\perp}')  + 
\nonumber\\ +
    \left[ 
    \frac{\overline m^2_{1} + \vec k_{\!\perp}^{\,2}}{x} +
    \frac{\overline m^2_{2} + \vec k_{\!\perp}^{\,2}}{1-x}  
    \right]
    \psi_{\lambda_1\lambda_2}(x,\vec k_{\!\perp})  
,\label{eq:9}\end{eqnarray} 
an integral equation with the kernel
\begin{eqnarray} \nonumber
    U_{\lambda_1\lambda_2;\lambda_1'\lambda_2'}
    (x,\vec k_{\!\perp};x',\vec k_{\!\perp}') = -
    \frac{4\overline m_1 \overline m_2}{3\pi^2} 
\\ \times
    \frac{\overline\alpha(Q)}{Q^2} \overline R(Q)
    \frac{S_{\lambda_1\lambda_2;
    \lambda_1'\lambda_2'}(x,\vec k_{\!\perp};x',\vec k_{\!\perp}')}
    {\sqrt{ x(1-x) x'(1-x')}}
.\label{eq:10}\end{eqnarray} 
Here, $M ^2$ is the eigenvalue of the invariant-mass squared. 
The associated eigenfunction $\psi\equiv\Psi_{q\bar q}$ 
is the probability amplitude 
$\langle x,\vec k_{\!\perp};\lambda_{1},\lambda_{2}\vert\psi\rangle$ 
for finding a quark with momentum fraction $x$, 
transversal momentum $\vec k_{\!\perp}$ 
and helicity $\lambda_{1}$,
and correspondingly the anti-quark with
$1-x$, $-\vec k_{\!\perp}$ and $\lambda_{2}$.
The (effective) quark masses $\overline m _1$  and $\overline m _2$ 
and the (effective) coupling constant $\overline\alpha$ are given below. 
The mean Feynman-momentum transfer of the quarks is
denoted by $Q^2=Q ^2 (x,\vec k_{\!\perp};x',\vec k_{\!\perp}')$, 
\begin{eqnarray}
   Q ^2 = 
   -\frac{1}{2}\left[(k_{1}-k_{1}')^2 + (k_{2}-k_{2}')^2\right]
,\end{eqnarray}
the spinor factor $S=S(x,\vec k_{\!\perp};x',\vec k_{\!\perp}')$ by 
\begin{eqnarray} 
 &&\hspace{-1.8em}S_{\lambda_1\lambda_2;\lambda_1'\lambda_2'}
   (x,\vec k_{\!\perp};x',\vec k_{\!\perp}')\!=\!
\label{eq:12}\\
&\hspace{-1em}=&\!\!\!\!\!\left[\overline u(k_1,\lambda_1)\gamma^\mu
   u(k_1',\lambda_1')\right] 
   \left[ \overline v(k_2',\lambda_2') \gamma_\mu 
   v(k_2,\lambda_2)\right] 
\nonumber\\
&\hspace{-1em}=&\!\!\!\!\!\left[\overline u(k_1,\lambda_1)\gamma^\mu
   u(k_1',\lambda_1')\right] 
   \left[\overline u (k_2,\lambda_2)\gamma_\mu
   u(k_2',\lambda_2')\right] 
,\nonumber\end{eqnarray}
since 
\(
   \overline v(k_2',\lambda_2') \gamma_\mu 
   v(k_2,\lambda_2) =
   \overline u (k_2,\lambda_2)\gamma_\mu
   u(k_2',\lambda_2')  
\)
holds as a general identity.
One deals thus only with the $u$-spinors.
Opposed to the earlier conventions \cite{BroPauPin98} 
they are normalized in Eq.(\ref{eq:12}):
\(
   \overline u (k,\lambda) u (k,\lambda') = 
   \delta_{\lambda\lambda'}
.\)
The spinor factor S is tabulated explicitly in \cite{Pau00}.
The regulator function $\overline R(x',\vec k'_{\!\perp};\Lambda)$ 
restricts the range of integration as function of some mass scale 
$\Lambda$.
Note that Eq.(\ref{eq:9}) is a fully relativistic equation.
I have derived essentially the same equation also with 
Wegner's Hamiltonian flow equations, see \cite{Pau00}.

\section{Renormalization}
The effective Hamiltonian in Eq.(\ref{eq:9}) depends on a
regulator scale $\Lambda$ through three quantities.\\ 
First, it depends on $\Lambda$ through the effective quark masses 
$\overline m _f \equiv \overline m _f (\Lambda)$
which are hidden also in the Dirac spinors. 
They are given in Eq.(90) of \cite{Pau98} in terms of the
bare $\alpha$ and $m_f$:
\begin{eqnarray}
   \overline m_f^2 &=& m_f^2 
   \left(1+\frac{\alpha}{\pi}\frac{n_c^2-1}{2n_c}
   \ln{\frac{\Lambda^2}{m_g^2}}\right)
.\label{eq:13}\end{eqnarray}
In the expression a second regularization parameter
$m_g$ appears which conceptually is a kinematical gluon mass.
The corresponding gluon mass diagram gives
\begin{eqnarray}
   \overline m_g^2 &=& m_g^2 - \frac{\alpha}{4\pi} 
   \sum_{f=1}^{n_f} m_f^2 \ln{\frac{\Lambda^2}{4m_f^2}}
,\label{eq:14}\end{eqnarray}
see Eq.(91) of \cite{Pau98}. 
The physical gluon mass must vanish due to gauge invarinace,
thus $\overline m_g=0$, which expresses $m_g$ in terms of $m_f$.\\
Second, it depends on $\Lambda$ through the effective coupling 
$\overline \alpha(Q) \equiv \overline \alpha(\Lambda,Q)$.
The expression in Eq.(100) of \cite{Pau98} is rewritten 
here conveniently in terms of an arbitrary scale $\kappa$:
\begin{eqnarray}
   \frac{1}{\overline{\alpha}(\Lambda,Q)} 
   &=& \frac{1}{\alpha} 
   - \frac{11n_c-2n_f}{12\pi}\ln{(\Lambda^2/\kappa^2)} 
\nonumber\\
   &+& \frac{11n_c}{12\pi} \ln{\left((\mu_g^2+Q^2)/\kappa^2\right)}
\nonumber\\
   &-& \frac{2}{12\pi} \sum_{f=1}^{n_f}
   \ln{\left((\mu_f^2+Q^2)/\kappa^2\right)}
,\label{eq:15}\end{eqnarray}
with $\mu_f = 2 m_f$ and $\mu_g = 2 m_g$.\\
Third, the Hamiltonian depends on $\Lambda$ through 
the regularization function which here is the soft cut-off
\begin{eqnarray}
   \overline R(Q) \equiv \overline R(\Lambda,Q) 
   = \frac{\Lambda^2}{\Lambda^2+Q^2}
.\end{eqnarray}
The dependence on the unphysical parameter $\Lambda$  
must be removed,
\begin{eqnarray}
   \frac{d}{d\Lambda} H_\mathit{LC} \left(
   \overline m(\Lambda), 
   \overline \alpha(\Lambda),
   \overline R(\Lambda)   \right) = 0 
,\end{eqnarray}
as required by renormalization theory, but how?
The non-perturbative renormalization of $H$ 
was stuck for many years by the fact that the vertex function 
$\overline \alpha(\Lambda)$ and the regulator $\overline R(\Lambda)$
are so intimately coupled in Eq.(\ref{eq:9}).
It was always clear that one could add non-local counter terms 
\cite{Wilson6}, but it was utterly unclear how to construct them. 
The progress comes from the recent work on the $\uparrow\downarrow$-model 
\cite{FredericoFP01,FrewerPF01}:
Adding to $\overline R(\Lambda,Q)$ a counterterm $C(\Lambda,Q)$ 
and requiring that the sum 
$ R(\Lambda,Q)=\overline R(\Lambda,Q)+C(\Lambda,Q)$ 
be independent of $\Lambda$, determines $C(\Lambda,Q)$. 
One remains with 
\begin{eqnarray}
   R(\Lambda,Q) = \overline R(\Lambda,Q)+C(\Lambda,Q) =
   \frac{\mu^2}{\mu^2+Q^2}
.\end{eqnarray}
In line with renormalization theory, one then can go to the
limit $\Lambda\longrightarrow\infty$ and
$\mu$ becomes a parameter of the theory.

The cut-off dependence in $\overline \alpha(\Lambda,Q)$,
Eq.(\ref{eq:15}), can then be removed by replacing the bare coupling 
constant $\alpha$ by the cut-off dependent running coupling constant 
$\alpha_\Lambda$, \textit{i.e.}
\begin{eqnarray}
   \alpha_\Lambda=
   \frac{6\pi}{11n_c-2n_f}\ \frac{1}{\ln{(\Lambda/\kappa)}} 
.\label{eq:19}\end{eqnarray}
The \emph{renormalized} vertex function, 
\begin{eqnarray}
   \frac{1}{\overline{\alpha}(Q)} 
   &=& \frac{11n_c}{12\pi} \ln{\left((\mu_g^2+Q^2)/\kappa^2\right)}
\nonumber\\
   &-& \frac{2}{12\pi} \sum_{f=1}^{n_f}
   \ln{\left((\mu_f^2+Q^2)/\kappa^2\right)}
,\label{eq:20}\end{eqnarray}
does not depend explicitly on $\Lambda$, and the scale $\kappa$ 
becomes an other parameter of the theory.

In completing renormalization for the masses,
Eqs.(\ref{eq:13}) and (\ref{eq:14}) are first rewritten for $n_c=3$ as
\begin{eqnarray}
   \overline m_f^2 &=&  
   m_f^2+\frac{8m_f^2}{3\pi}
   \left(1-\frac{\ln{m_g/\kappa}}{\ln{\Lambda/\kappa}}\right)
   \alpha \ln{\frac{\Lambda}{\kappa}}
,\\
   m_g^2  &=&   
   \sum_{f=1}^{n_f} \frac{m_f^2 }{2\pi}
   \left(1-\frac{\ln{2m_f/\kappa}}{\ln{\Lambda/\kappa}}\right)
   \alpha \ln{\frac{\Lambda}{\kappa}}   
.\end{eqnarray}
Inserting the running coupling constant from Eq.(\ref{eq:19})
leaves us with 
\begin{eqnarray*}
\begin{array} {@{}l @{\,}c@{\,} l@{}}
   \displaystyle
   \overline m_f^2 &=&  
   \displaystyle
   m_f^2+\frac{8m_f^2}{3\pi} \frac{6\pi}{33-2n_f}
   \left(1-\frac{\ln{m_g/\kappa}}{\ln{\Lambda/\kappa}}\right) ,
\\
   \displaystyle
   m_g^2  &=&   
   \displaystyle
   \sum_{f=1}^{n_f} \frac{m_f^2 }{2\pi}\frac{6\pi}{33-2n_f}
   \left(1-\frac{\ln{2m_f/\kappa}}{\ln{\Lambda/\kappa}}\right) .
\end{array}\hspace{-5em}
\end{eqnarray*}
Finally, I go to the limit $\Lambda\rightarrow\infty$  
and express the bare masses in terms of the dressed ones:
\begin{eqnarray}
   m_f^2 = \frac{33-2n_f}{49-2n_f} \overline m_f^2 
,\ m_g^2 = \frac{3}{49-2n_f} \sum_{f=1}^{n_f}\overline m_f^2
.\end{eqnarray}
This completes the program of renormalization:
For the first time, ever, the dependence on a cut-off $\Lambda$
has been removed completely from a field theoretical Hamiltonian.
Notice that this step rests on the contributions
\cite{FredericoFP01,FrewerPF01} to this meeting.

\begin{figure} [t]
  \scalebox{0.46}{\includegraphics{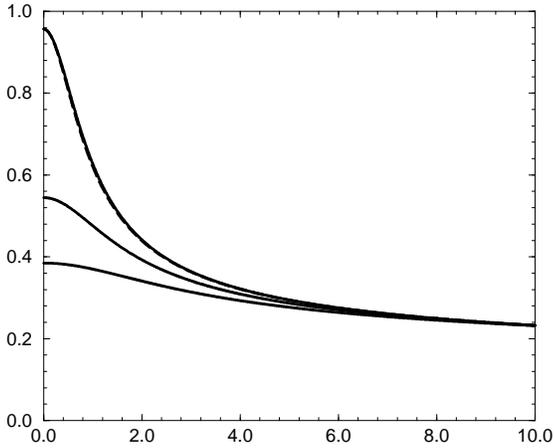}}
  \caption{\label{fig:4}  
     The vertex function $\overline \alpha(Q)$ versus $Q$ in GeV, 
     with for different flavor numbers $n_f=4,5,6$ (top to bottom);
     all $\overline m_f=350$~MeV.} 
\end{figure}

\begin{figure} [t]
  \scalebox{0.46}{\includegraphics{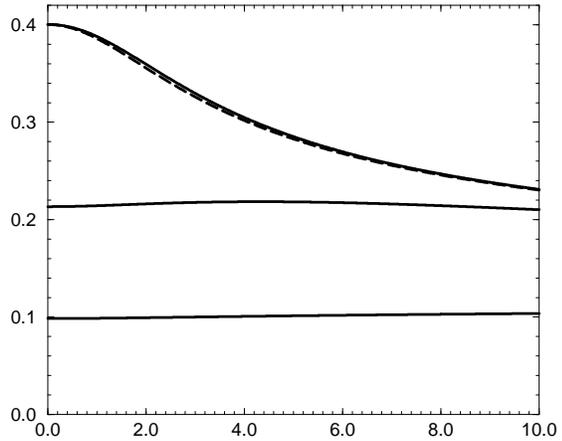}}
  \caption{\label{fig:5}  
     The vertex function $\overline \alpha(Q)$ versus $Q$ in GeV, 
     with for different flavor numbers $n_f=4,5,6$ (top to bottom).} 
\end{figure}

%
\section{The locking of the coupling constant}
The Lagragian for QCD has 7 parameters:
the 6 flavor quark masses $m_f$ and the coupling constant $\alpha$.
The renormalized effective Hamiltonian has one parameter more:
The 6 flavor masses $\overline m_f$, and the two scales $\kappa$ and $\mu$.
This is in full accord with renormalization theory,
since whatever the model, one has a scale at which one experiments.

The renormalized vertex function of Eq.(\ref{eq:20}) deserves some further
discussion. Most importantly it has a \emph{finite value at $Q=0$}.
The coupling constant locks its-self, as one says.

One should think that $\kappa$ is entirely fixed by the coupling constant
of measured at sufficiently high $Q$. But taking, as usual, the value 
$\overline \alpha(M_Z)=0.118$ at the Z mass $M_Z=91.2\mbox{ GeV}$,
one observes a rather dramatic dependence of $\kappa$
on the number of flavors included:
\begin{eqnarray}
\begin{array} {@{}ccccc@{}}
   n_f & \kappa     & \alpha_0 & \mu_g      & \mu_b       \\
    -  &[\mbox{MeV}]&    -     &[\mbox{MeV}]&[\mbox{MeV}] \\
    4  & 153.1      & 0.9566   & 387.7      & 336.7 \\
    5  & 87.84      & 0.5446   & 434.1      & 359.6 \\
    6  & 45.33      & 0.3848   & 488.2      & 467.2  
\end{array}
\label{eq:24}\end{eqnarray}
Changing $n_f$ from 4 to 6 changes $\kappa$ by a factor of 4!
The dependence of $\alpha_0 \equiv \overline \alpha(0)$ on $n_f$ is
less pronounced even if one puts all flavor masses equal to
$\overline m_f = 350 \mbox{ MeV}$, as done conveniently
in Eq.(\ref{eq:24}). 
The corresponding functions $\overline \alpha (Q)$ are displayed 
in Figure~\ref{fig:4}. 
The $n_f+1$ parameters in Eq.(\ref{eq:20}) are unpleasant to
work with and it is useful to introduce the approximate expression
\begin{eqnarray}
   \overline \alpha _b(Q)  
   &=& \frac{12\pi}{33-2n_f} 
   \frac{1}{\ln{\left((\mu_b^2+Q^2)/\kappa^2\right)}}
.\end{eqnarray}
The only parameter $\mu_b$ is fixed by $\alpha _0$ and given 
in Eq.(\ref{eq:24}) as well. As shown in Figures~\ref{fig:4}
and \ref{fig:5} by the dashed line, 
$\overline \alpha _b(Q)$ is almost 
un-discernible from $\overline \alpha (Q)$.

Using the more physical mass parameters from Eq.(\ref{eq:31})
produces 
\begin{eqnarray}
\begin{array} {@{}ccccc@{}}
   n_f & \kappa     & \alpha_0 & \mu_g      & \mu_b       \\
    -  &[\mbox{MeV}]&    -     &[\mbox{GeV}]&[\mbox{MeV}] \\
    4  & 153.1      & 0.4002   & .9941      & 1.007 \\
    5  & 87.88      & 0.2133   & 2.981      & 4.099 \\
    6  & 75.23      & .09844   & 99.14      & 685.8  
\end{array}
\end{eqnarray}
and the corresponding curves $\overline \alpha (Q)$ in Figure~\ref{fig:5}.
Latest here I have to abandon my earlier conjecture \cite{Pau98}
that a momentum-dependent vertex function could be related
to confinement in any way. In fact, the above curves have so little 
structure that one can replace them in a bound state calculation
by the constant $\alpha_0$.
Henceforward I will give up thus $\kappa$ in favor of $\alpha=\alpha_0$ 
and change notation from $\overline m_f$ to $m_f$ .

\section{The $\uparrow\downarrow$-model as an application}
%
\begin{table} [t]
\caption{\label{tab:2.2}  
   The calculated mass eigenvalues in MeV. 
   Those for singlet-1s states are given in the lower,
   those for singlet-2s states in the upper triangle.}
\begin{tabular}{c|rrrrrr} 
     & $\overline u$ & $\overline d$ 
     & $\overline s$ & $\overline c$ & $\overline b$ \\ \hline
    $u$ &         &      768&      871&     2030&     5418 \\
    $d$ &      140&         &      871&     2030&     5418 \\
    $s$ &      494&      494&         &     2124&     5510 \\
    $c$ &     1865&     1865&     1929&         &     6580 \\
    $b$ &     5278&     5278&     5338&     6114&          
\end{tabular}
\end{table}

In light-cone parametrization, the quarks are at relative rest
when $\vec k _{\!\perp}= 0$ and 
$ x = \overline x \equiv m_1/(m_1+m_2)$.
For very small deviations from these equilibrium values 
the spinor matrix is proportional to the unit matrix, with \cite{Pau00}
\begin{eqnarray}
   \langle\lambda_1,\lambda_2\vert S\vert\lambda_1'\lambda_2'\rangle
   \sim 4 m_1 m_2 
   \ \delta_{\lambda_1,\lambda_1'}
   \ \delta_{\lambda_2,\lambda_2'}
.\end{eqnarray}
For very large deviations from equilibrium, particularly for
$\vec k_{\!\perp}^{\prime\,2} \gg \vec k_{\!\perp} ^{\,2}$,
holds  
\begin{eqnarray}
   Q ^2 \simeq\vec k_{\!\perp}^{\prime\,2} 
,\hskip2em\mbox{ and } \hskip2em
   \langle\uparrow\downarrow\vert S\vert\uparrow\downarrow\rangle 
   \simeq 2\vec k_{\!\perp}^{\prime\,2}
.\end{eqnarray}
Both extremes are combined in the $\uparrow\downarrow$-model \cite{Pau00}:
\begin{eqnarray}
 &&\frac{S} {Q ^2} \equiv 
   \frac{4 m_1 m_2}{Q ^2} + 2 
   \Longrightarrow
   \frac{4 m_1 m_2}{Q ^2} + 2 R(\Lambda,Q) 
,\nonumber\\
 &&\mathrm{with}\quad 
   R(\Lambda,Q) = \frac{\mu^2}{\mu^2+Q^2}
.\end{eqnarray} 
It interpolates between two extremes:
For small momentum transfer, the `2' generated by the hyperfine interaction
is unimportant and the dominant Coulomb aspects of the first term prevail.
For large momentum transfers the Coulomb aspects are
unimportant and the 2
dominates.
Eq.(\ref{eq:9}) therefore is replaced by
\begin{eqnarray} 
    \lefteqn{ 
    M^2
    \psi(x,\vec k_{\!\perp}) = \left[ 
    \frac{ m^2_{1} + \vec k_{\!\perp}^{\,2}}{x} +
    \frac{ m^2_{2} + \vec k_{\!\perp}^{\,2}}{1-x}  
    \right]\psi(x,\vec k_{\!\perp}) }
\nonumber\\  
    \lefteqn{ 
    -
    \frac{\alpha}{3\pi^2} \!\int 
    \frac{ dx' d^2 \vec k_{\!\perp}'}
    {\sqrt{ x(1-x) x'(1-x')}}\ \psi(x',\vec k_{\!\perp}')\ \times}
\nonumber\\  & & \ \times
    \left(\frac{4 m_1 m_2}{Q ^2} + 
    \frac{2\mu^2}{\mu^2+Q^2}\right),  
\label{eq:30}\end{eqnarray} 
where 
$\psi(x,\vec k_{\!\perp})\equiv
 \langle x,\vec k_{\!\perp}; \uparrow,\downarrow
 \vert \psi\rangle $.
With the  canonical 8 parameters of the 
$\uparrow\downarrow$-model \cite{Pau00},
\begin{eqnarray}
\begin{array} {@{}c@{\,}c@{\quad}c@{\,}c@{\,}c@{\ \ }c@{\,}c@{\,}c@{}}
   \alpha & \mu  & m_u & m_d & m_s &  m_c &  m_b & m_t  \\
    0.690 & 1.33 & 406 & 406 & 508 & 1.67 & 5.05 & 174  \\
      -   &\mbox{GeV}&\mbox{MeV}&\mbox{MeV}&\mbox{MeV}
                     &\mbox{GeV}&\mbox{GeV}&\mbox{GeV},
\end{array}
\label{eq:31}
\end{eqnarray}
all masses of the physical mesons have been calculated
according to Eq.(\ref{eq:30}). 
They are compiled in Table~\ref{tab:2.2}.
The empirical masses are compiled in Table~\ref{tab:1.2}.
The agreement between the two is amazing.
To the best of my knowledge there is no other model which can describe
\emph{ all mesons} quantitatively from the $\pi$ up to the $\Upsilon$
from a common point of view, which here is QCD. 

The proposed pion of the $\uparrow\downarrow$-model is rather
different from the pions in the literature.
I have found no evidence that the vacuum condensates are important,
but I conclude that the pion is describable by a QCD-inspired theory:
The very large coupling constant in conjunction 
with a very strong hyperfine interaction
makes it a ultra strongly bounded system of constituent quarks.
More then 80 percent of the constituent quark mass is eaten up by 
binding effects.
No other physical system has such a property.

The numerical wavefunction $\psi(x,\vec k_{\!\perp})$ 
can be fitted with only one free parameter, \textit{i.e.}
\begin{eqnarray}
   &&\psi(x,\vec k _{\!\perp}) =
   \frac{\mathcal{N}}{\sqrt{x(1-x)}} 
\nonumber\\ &&\times
   \frac
   {\left(1+\displaystyle
   \frac{m^2\left(2x-1\right)^2+\vec k_{\!\perp}^{\,2}}
   {4x(1-x)\ m^2} \right)^{\frac{1}{2}}}
   {\left(1+\displaystyle
   \frac{m^2\left(2x-1\right)^2+\vec k_{\!\perp}^{\,2}}
   {4x(1-x)\ p_a^2} \right)^2} 
,\label{eq:25}\end{eqnarray}
with $p_a=1.338 m$ \cite{Pau00}.
The explicite form of the wavefunction can used to calculate the form factor 
and thus the exact root-mean-square radius 
$\langle r^2\rangle = -6\left.{dF_2(Q^2)}/{dQ^2}\right\vert_{Q^2=0}$ 
analytically \cite{PaM01}.  
The size of the $q\bar q$  wavefunction turns out as
$\langle r^2\rangle = (0.33\mbox{ fm})^2$, 
half as large as the empirical value   
$\langle r^2\rangle_{\mathrm{exp}} = (0.67\mbox{ fm})^2$.

\begin{table} [t]
\caption{\label{tab:1.2}  
   Empirical masses of the flavor-off-diagonal physical mesons in MeV.
   Vector mesons are given in the upper, scalar mesons
   in the lower triangle.}
\begin{tabular}{c|rrrrrr} 
     & $\overline u$ & $\overline d$ 
     & $\overline s$ & $\overline c$ & $\overline b$ \\ \hline
 $u$ &      & 768  & 892  & 2007 & 5325 \\ 
 $d$ & 140  &      & 896  & 2010 & 5325 \\ 
 $s$ & 494  & 498  &      & 2110 &  --- \\ 
 $c$ & 1865 & 1869 & 1969 &      &  --- \\ 
 $b$ & 5278 & 5279 & 5375 &  --- &      \\ 
\end{tabular}
\end{table}

The parameter $p_a=1.338\,m$ in Eq.(\ref{eq:25}) plays the role of an
effective Bohr momentum of the constituents in the pion.
The mean momentum of the constituents is thus 40 percent 
larger than their mass, which means that 
they move highly relativistically quite in contrast to the 
constituents of atoms or nuclei.
No wonder that potential models thus far have failed for the pion.

This completes one of my goals: I have a pion with the correct mass, 
and I have an analytic expression for its light-cone wave function.
Eq.(\ref{eq:25}) could be used thus as a baseline for calculating 
the higher Fock-space amplitudes, as explained in \cite{Pau98}.
It could well be that the wavefunction obtained from such a  
simple model suffices already to be consistent with 
recent experiments \cite{Ash00}.

\section{Discussion: Front or instant form?}
Eq.(\ref{eq:9}) is a frame frame-independent,
covariant, and fully relativistic front-form equation,
with certain boosts being kinematic and trivial \cite{BroPauPin98}.
One pays for these advantages with the fact 
that the transversal components for total angular momentum
$\vec J = \vec L + \vec S$ 
(the spin $\vec S=\frac{1}{2}(\sigma_1+\sigma_2)$ is not 
to be confused with the spinor factors $S$)
are complicated dynamical operators in the front form, 
see for example \cite{BroPauPin98}.
Only $J_z$ is simple and kinematic.
The eigenvalues and eigenfunctions of Eq.(\ref{eq:9}) 
can thus \underline{not} be labeled with $J$.
Despite this, Trittmann and Pauli \cite{TriPau00}, 
in their numerical solution of the QED-version
of Eq.(\ref{eq:9}) for different $J_z$, 
have done so by using
the standard (non-relativistic) spectroscopic 
terms $^{2S+1} L _{J} ^{J_z}$.
By inspection of the numerical results they found
that the eigenvalues can be arranged in  
multiplets which are ($2J+1$)-fold degenerate modulo
numerical accuracy.
The authors could not find a plausible answer for that 
in terms of the light-cone formalism.

Now, we seem understand that better.
In the contribution to this session,
Krassnigg \cite{Krassnigg01} shows that there exists a unitary
tranformation $\Omega$ which transforms the typical 
combination of Lepage-Brodsky spinors in Eq.(\ref{eq:9}) 
to an other combination with only Bj\o rken-Drell spinors.

Unitary transfomartions do not change the eigenvalue
and Eq.(\ref{eq:9}) is identically transcribed 
to an equation for the reduced wave function $\varphi_{s_1 s_2}(\vec k)$,
which reads for equal masses 
\begin{eqnarray}
 \begin{array}{l@{}l@{}l@{}l@{}} 
    &&\left[M^2-4\left(m^2+\vec k^{\,2}\right)\right] 
    \varphi _{s_1 s_2}(\vec k) 
\label{eq:33}\\ 
    &=&
    \displaystyle 
    \sum _{s'_1, s'_2} 
    \!\int\!\! d^3 \vec k' 
    \ \widetilde U _{s_1 s_2;s'_1 s'_2} (\vec k;\vec k')
    \ \varphi_{s'_1 s'_2}(\vec k') .  
 \end{array}  
\hspace{-6em}\end{eqnarray} 
Since the spinors $u(k,s)$ in the kernel
\begin{eqnarray}
 \begin{array}{l@{}l@{}l@{}l@{}} 
    &\ &
    \displaystyle 
    \widetilde U _{s_1 s_2 ;s'_1 s'_2} = 
    -\frac{8m}{3\pi^2}
    \frac{\overline\alpha(Q)}{Q^2}R(Q) \times
\\ &&
   \displaystyle 
   \left[\overline u(k_1,s_1)\gamma^\mu
   u(k_1',s_1')\right] 
   \left[\overline u (k_2,s_2)\gamma_\mu
   u(k_2',s_2')\right] , 
 \end{array}  
\hspace{-6em}\end{eqnarray} 
by definition are Bj\o rken-Drell spinors, 
one can not recognize the front-form orign 
of this equation, see \cite{Krassnigg01} for further details.
They are a set of four coupled integral equations
in the usual momentum space. 
Formally spoken, they are instant-form
equations in the rest frame, and, by inspection, 
they are invariant under spatial rotations.
Its eigenfunctions can therefore be labeled as
$\varphi_{J(L)S}$, as usual,
with eigenvalues $M_{J(L)S}^2$ being 
($2J+1$)-fold strictly degenerate multiplets.

These aspects can also be reversed.
Suppose that some phenomenological model yields
momentum space wavefunctions $\varphi_{s_1 s_2}(\vec k)$.
The transformations in \cite{Krassnigg01}, 
which lead to Eq.(\ref{eq:33}), can be inverted and 
used to generate light-cone wave functions
$\psi_{\lambda_1\lambda_2}(x,\vec k _{\!\perp})$
with helicities $\lambda_1$ and $\lambda_2$.
These can be used then as a reasonable approximation 
in existing formulas for the cross-sections.

\section{Perspectives}
The light-cone community has not yet solved its homework problem,
but it has gone a long way:
(1) The role of zero modes and vacuum structure is better understood.
(2) Effective interactions can be formulated and even renormalized.
(3) Bound-state wavefunctions can be calculated with a technical 
    effort comparable to or less than in the instant form.
(4) Simple models can generated which are not in conflict with experiments.
(5) Last not least, much of the work can be done analytically.\\
Despite the limited progress one should not be discouraged 
from continued efforts on the homework problem.
After all, the Hamiltonian approach to \emph{any} field theory an in
\emph{any} form has been disrupted in 1949 
when Feynmans action oriented approach did appear on the scene.

Not solved are the aspects of confinement.
At present one does not understand its orign.
Not solved are also the aspects of the chiral phase transition.
But one should emphasize that the solution to the bound-state
problem takes place at temperature zero, 
possibly after a phase transition.
Due to the fit to experiment, 
quark masses are finite and actually large.
The present approach thus can not contribute to question like
``What happens if quark masses vanish?''
It starts where other approaches end.

In QED, hyperfine interactions and the Lamb shift are comparable in size.
For QCD, the importance of the hyperfine interaction has
been quantified by the $\uparrow\downarrow$-model,
but the Lamb shift is a completely open question. 

In QED, the Lamb shift arises by a photon in flight moving 
relative to the  hydrogen bound-state.
That alone suffices to give part of the answer for QCD, by Eq.(\ref{eq:8}).
Going to the diagonal representation gives
\begin{eqnarray}
   U_\mathit{Lamb} = VG_3V =
   \sum_n V\vert n\rangle\frac{1}{\omega-M_n^2 }\langle n\vert V
\end{eqnarray}
The symbol $\sum_n $ refers to a summation (or integration) over all
meson states and gluon states.
The invariant mass squared
\begin{eqnarray}
   M_n^2 = \frac{M_b^2 +q_{\!\perp}^2}{1-y} + \frac{q_{\!\perp}^2}{y} 
\end{eqnarray}
corresponds to a free colored gluon with longitudinal momentum 
fraction $y$ and transversal momentum $\vec q_{\!\perp}$.
It moves back to back to a colored meson bound-state of mass $M_b$.
We have not much of an idea on the bound-state spectrum of colored 
mesons, neither experimentally nor theoretically. 
We dont even know, whether such mesons are bound at all,
but I see no immediate objection why one could not give it a try
by the above methods, particularly a suitably adjusted
$\uparrow\downarrow$-model.

I conclude that the calculation of the Lamb shift
in QCD is an interesting and important problem
particularly for the pion. 
Can we challenge the lattice community
to get help on that?


\begin{thebibliography}{99}

\bibitem{Leutwyler01}
     H. Leutwyler, 
     these  proceedings.

\bibitem{Roberts01}
     C. Roberts, \textit{l.c.}

\bibitem{FredericoFP01}
     T. Frederico \textit{et al.}, 
     \textit{l.c.},
     hep-ph/0111136 
 
\bibitem{Schweiger01}
     A. Schweiger, \textit{l.c.}

\bibitem{Plessas01}
     W. Plessas, 
     \textit{l.c.}

\bibitem{Krassnigg01}
     A. Krassnigg \textit{et al.}, \textit{l.c.},
     hep-ph/0111260



\bibitem{Mangin01}
     M. Mangin-Brinet, \textit{l.c.}
 
\bibitem{FrewerPF01}
     M. Frewer \textit{et al.}, 
     hep-ph/0111039 

\bibitem{Walhout01}
     T. Walhout, \textit{l.c.}

\bibitem{Karmanov01}
     T. Karmanov, \textit{l.c.}

\bibitem{Ligterink01}
     R. Ligterink, \textit{l.c.}

\bibitem{Sugihara01}
     T. Sugihara, \textit{l.c.}

\bibitem{vanIersel01}
     M. van Iersel, \textit{l.c.}

\bibitem{Ash00}
     D. Ashery, 
     Nucl. Phys. B (Proc.Suppl.) \textbf{90} (2000) 67;
     see also \textit{l.c.}

\bibitem{Wegner00} 
     Franz Wegner, Nucl. Phys. B (Proc.Suppl.) \textbf{90} (2000) 141;
     H.C. Pauli, Nucl. Phys. B (Proc.Suppl.) \textbf{90}(2000) 147.

\bibitem{Hil00}
     J. Hiller, Nucl. Phys. B (Proc.Suppl.) \textbf{90} (2000) 170,
     and \textit{l.c.}

\bibitem{Schierholz00}
     G. Schierholz, 
     Nucl. Phys. B (Proc.Suppl.) \textbf{90} (2000) 207.

\bibitem{TriPau00} 
     U. Trittmann and H.C. Pauli, 
     Nucl. Phys. B (Proc.Suppl.) \textbf{90} (2000) 161.

\bibitem{Pau00} 
     H.C. Pauli, Nucl. Phys. B (Proc.Suppl.) \textbf{90} (2000) 147, 259.




\bibitem {pab85a}  H.C. Pauli and S.J. Brodsky,      
   Phys. Rev. {\bf D32} (1985) 1993, 2001.

\bibitem{BroPauPin98} 
     S.J. Brodsky, H.C. Pauli, and S.S. Pinsky, 
     Phys. Rep. \textbf{301} (1998) 299-486. 

\bibitem{Schilling2000}
     K. Schilling,
     Nucl. Phys. B(Proc.Suppl.) \textbf{83} (2000) 140.

\bibitem{Wilson6} K.G.~Wilson,  T.S.~Walhout,  A.~Harindranath, 
     W.M.~Zhang, R.J.~Perry,  S.D.~G{\l}azek, 
     Phys. Rev. \textbf{D49} (1994) 6720-6766. 

\bibitem{GlazekWilson12}
     S.D. G{\l}azek and K.G. Wilson, 
     Phys. Rev. \textbf{D48} (1993) 5863;
     \textit{ibid.} \textbf{D49} (1994) 4214. 

\bibitem{Weg94} 
     F. Wegner, Ann. Physik \textbf{3} (1994) 77. 

\bibitem{Pau98}
     H.C. Pauli,
     Eur. Phys. J. \textbf{C7} (1998) 289.
     hep-th/9809005,
     and
%
     in: New directions in Quantum Chromodynamics,
     C.R. Ji and D.P. Min, Eds.,
     American Institute of Physics, 1999, p. 80-139.
     hep-ph/9910203.

\bibitem{PaM01}
     H.C. Pauli, and A. Mukherjee,
     to appear in Int.Jour.Mod.Phys. (2001),
     hep-ph/0104175.
     H.C. Pauli,
     Submitted to Nucl.Phys. A(2001).

\end{thebibliography}
\end{document}